\documentclass[structabstract]{aa}  
\usepackage{graphicx}
\usepackage{subfig}
\usepackage{nth}
\usepackage{txfonts}
\usepackage{natbib}
\bibpunct{(}{)}{,}{a}{}{,}
\newcommand{\ms}{\mbox{${\rm m\,s}^{-1}$}}
\newcommand{\msd}{\mbox{${\rm m\,s}^{-1}\,d^{1/3}$}}
\newcommand{\Msolar}{\mbox{${M}_{\sun}$}}
\newcommand{\Rsolar}{\mbox{${R}_{\sun}$}}

\newcommand{\Mjup}{\mbox{${M}_{J}$}}

\newcommand{\rhosun}{\mbox{$\rho_{\sun}$}}
\newcommand{\Rjup}{\mbox{${R}_{J}$}}
\newcommand{\rhojup}{\mbox{$\rho_{J}$}}
\newcommand{\mym}{\mbox{$\muup$}\rm m}

\newcommand\T{\rule{0pt}{2.2ex}}

\begin{document}
   \title{A photometric study of the hot exoplanet WASP-19b\thanks{Based on photometric observations made with HAWK-I on the ESO VLT/UT4 (Prog. ID 084.C-0532),
 EulerCam on the Euler-Swiss telescope and the Belgian TRAPPIST telescope, as well as archive data from the Faulkes South Telescope, CORALIE on the Euler-Swiss telescope, 
HARPS on the ESO 3.6~m telescope (Prog. ID 084-C-0185), and HAWK-I (Prog. ID 083.C-0377(A)).}$^{,}$\thanks{The photometric time series data in this work are only available in electronic form
at the CDS via anonymous ftp to cdsarc.u-strasbg.fr (130.79.128.5) or via http://cdsweb.u-strasbg.fr/cgi-bin/qcat?J/A+A/}
}

   \author{M.~Lendl
          \inst{1}
          \and
          M.~Gillon
          \inst{2}
          \and
          D.~Queloz
          \inst{1} 
          \and
          R.~Alonso
          \inst{3,4}
          \and
          A.~Fumel
          \inst{2}
          \and
          E.~Jehin
          \inst{2}
          \and
          D.~Naef
          \inst{1}
}

   \institute{Observatoire de Gen\`eve, Universit\'e de Gen\`eve, Chemin des maillettes 51, 1290 Sauverny, Switzerland,
              \email{monika.lendl@unige.ch}
         \and
             Universit\'e de Li\`ege, All\'ee du 6 ao\^ut 17, Sart Tilman, Li\`ege 1, Belgium
         \and
             Instituto de Astrof\'isica de Canarias, C/V\'ia Lact\'ea s/n, 38205, La Laguna, Spain
         \and
	     Universidad de La Laguna, Departamento de Astrof\'isica, 38200, La Laguna, Spain
}

   \date{Received: 14 December 2012, Accepted: 27 January 2013, Published: 13 March 2013 }

  \abstract
   {The sample of hot Jupiters that have been studied in great detail is still growing. In particular, when the planet transits
its host star, it is possible to measure the planetary radius and the planet mass (with radial velocity data). 
For the study of planetary atmospheres, it is essential to obtain transit and occultation measurements at multiple wavelengths.}
   {We aim to characterize the transiting hot Jupiter WASP-19b by deriving accurate and precise planetary parameters from a
dedicated observing campaign of transits and occultations.}
   {We have obtained a total of 14 transit lightcurves in the \textit{r'-Gunn}, \textit{I-Cousins}, \textit{z'-Gunn}, and \textit{I+z'} filters and 10 occultation lightcurves
 in \textit{z'-Gunn} using EulerCam on the Euler-Swiss telescope and TRAPPIST. We also obtained one lightcurve through the narrow-band
NB1190 filter of HAWK-I on the VLT measuring an occultation at 1.19 {\mym}. We performed a global MCMC analysis of all new data, 
together with some archive data in order to refine the planetary parameters and to measure the occultation depths in z'-~band and at 1.19 {\mym}.}
   {We measure a planetary radius of $R_{p} = 1.376\pm0.046$ {\Rjup}, a planetary mass of $ M_{p} = 1.165\pm0.068$ {\Mjup}, and find a 
very low eccentricity of $e = 0.0077_{-0.0032}^{+0.0068}$, compatible with a circular orbit. 
We have detected the z'-band occultation at $3 \sigma$ significance and measure it to be $\delta F_{occ,z'} = 352\pm116$~ppm, more than a factor of 2
smaller than previously published.
The occultation at 1.19 {\mym} is only marginally constrained at $\delta F_{occ,NB1190} = 1711_{-726}^{+745}$~ppm.}
   {We show that the detection of occultations in the visible range is within reach, even for 1m class telescopes if a considerable number of 
individual events are observed. Our results suggest an oxygen-dominated atmosphere of WASP-19b, making the planet an interesting test case for 
oxygen-rich planets without temperature inversion.}


  \maketitle

\section{Introduction}
\label{sec:int}

At the time of writing, about 290 planets have been confirmed as transiting in front of their parent stars\footnote{Based on 
www.exoplanet.eu \citep{Schneider11}}. Via the precise measurement of transit lightcurves, we are able to constrain the planetary
radius, orbital inclination, and mass (usually with the help of radial velocity measurements), hence the planetary density. 

Transiting planets open up a window onto the study of planetary atmospheres, their structure and composition. 
High-precision spectroscopic or spectro-photometric observations of planetary transits allow us to search for wavelength 
dependencies in the effective planetary radius, and from there conclude on the molecular species present in the planetary 
atmosphere. Also, from the transit lightcurve an independent measurement of the stellar mean density can be obtained
\citep{Seager03}, which is particularly useful since it can be used to refine the host stars' parameters, namely its
radius and mass. Having more accurate knowledge of the stellar parameters directly translates into more accurate physical
values for the planetary mass and radius. This has led to several campaigns that collect transit lightcurves of published
planets, as done by, e.g., \citet{Holman06} and \citet{Southworth09}.
Because the parameter measured from transit lightcurves is not the planetary radius itself but, in fact, the dimming of the star by the
planetary disk, these measurements are affected by the brightness distribution along the stellar disk, i.e. stellar limb darkening as well as occulted and 
nonocculted spots. Occulted dark spots or bright faculae lead to short-term flux variations during the transit as the planet passes areas
of the star having a different temperature and thus a different brightness. 
Nonocculted spots alter the stellar brightness outside of the planet's path, leading to a slight increase in the observed transit depth.
Depending on the spot distribution on the stellar surface, spots cause a rotational modulation of the stellar flux, with typical amplitudes 
of a few percent in the optical and timescales of several days. While the effect on the transit depth is weak (100 ppm for a 
typical brightness variation and transit depth of both 1 \%), 
it is within the precision needed to detect of elements through transmission spectroscopy. 
Next to these physical effects, ground-based photometric lightcurves are known to suffer from correlated noise due to airmass, seeing, 
or other external variations \citep{Pont06}. These effects can be mitigated by choosing optimal observation strategies (such as staying 
on the same pixels during the whole observation and defocusing in order to improve the sampling of the PSF) but can rarely be completely prevented. 
In this work, we collect a large number of transit lightcurves from 1m class telescopes and combine them to find not 
only a very precise but also an accurate measurement of the overall transit shape.

Observing the occultation of a transiting planet \citep{Charbonneau05,Deming05} allows us to measure the brightness ratio 
of planet and star and thus measure the flux emitted or, at shorter wavelengths reflected, by the planet.
At optical wavelengths, occultations have been measured almost exclusively from space \citep{Alonso09, Snellen09, Borucky09} using the
\textit{CoRoT} and \textit{Kepler} satellites. There have been few ground based observations \citep{Sing09, Lopez10, Smith11a}, and so far none 
of the detections has been independently confirmed.
Observations of occultations in the infrared have been plentiful, both from space \citep[starting with][]{Charbonneau05,Deming05}
and from the ground, e.g., \citet{DeMooij09,Gillon09a}, and \citet{Croll10}.
These observations provide information on the composition and the temperature profile of the planetary atmospheres.
Some planets, such as HD209458, show a temperature inversion at high altitudes \citep{Knutson08}, which is usually attributed to high abundances of 
TiO and VO \citep{Hubeny03,Fortney08}. These molecules are efficient absorbers of the stellar radiation heating up the high altitude atmosphere. 
However, it is not yet clear why some planets show inversions while others do not. As the number of planets with characterized atmospheres 
increases, the presence of inversions is turning out not to depend only on either the incident stellar flux \citep{Fortney08}
or the host star activity level \citep{Knutson10}. \citet{Spiegel09} argue that TiO might be depleted in many hot Jupiters by condensation 
and subsequent gravitational settling. Recently, \citet{Madhu11a} have suggested an additional connection between the C/O ratio 
and the presence of an inversion, because in atmospheres dominated by carbon, the main absorbers TiO and VO are not abundant enough to cause an inversion.
It is essential to increase the sample of well-studied transiting hot Jupiters and to provide accurate values for 
the measured occultation depths.

WASP-19b has been identified as a hot Jupiter by \citet{Hebb10}, based on data taken by the WASP survey \citep{Pollacco06}.
The slightly bloated ($\rho = 0.44$~$\rho_{J}$) planet with a mass near that of Jupiter ($M_{P} = 1.17 M_{J}$) 
is orbiting an $m_V = 12.3 $ G8V dwarf with a period of $0.79$ days. At this close separation, 
the planet is assumed to have been undergoing orbital decay moving it to its current orbital position at 1.21 times the Roche limit 
\citep{Hebb10,Hellier11a}. The star is known to be active showing a rotational modulation with a period of 10.5 days in the discovery lightcurves \citep{Hebb10}.
Also, anomalies in transit lightcurves attributed to spot crossings have been reported by \citet{Tregloan12}. 
The projected stellar rotation axis of WASP-19 is aligned with the planet's orbit \citep{Hellier11a, Albrecht12,Tregloan12}. 

Occultations of WASP-19b have been measured in the past by \citet{Gibson10} and 
\citet{Anderson10b} using HAWK-I in the K and H bands, respectively, as well as by \citet{Anderson11b} using the Spitzer Space Telescope at 3.6, 4.5, 5.8, and 8.0 {\mym}.  
Recently, \citet{Burton12} have published a z'-band lightcurve obtained with ULTRACAM during one occultation of WASP-19b, 
claiming its detection at $880 \pm 119$~ppm. 
From the ensemble of measurements, \citet{Anderson11b} and \citet{Madhu12a} determine that WASP-19b does not possess a temperature inversion.
Using the z'-band value of \citet{Burton12}, models favor a C-rich atmosphere.
For the eccentricity of WASP-19b, \citet{Anderson11b} derive a 3-$\sigma$ upper limit of $e < 0.027$.

In this paper we present results from an intense observing campaign of transits and occultations of WASP-19 obtained in both the optical and IR light.
We describe all observations and their reduction in Section \ref{sec:obs} and give details on the modeling
in Section \ref{sec:mod}. In Sections \ref{sec:res} and \ref{sec:dis} we present and discuss the results before concluding in Section \ref{sec:con}.

\begin{figure*}
\includegraphics[width=0.9\linewidth]{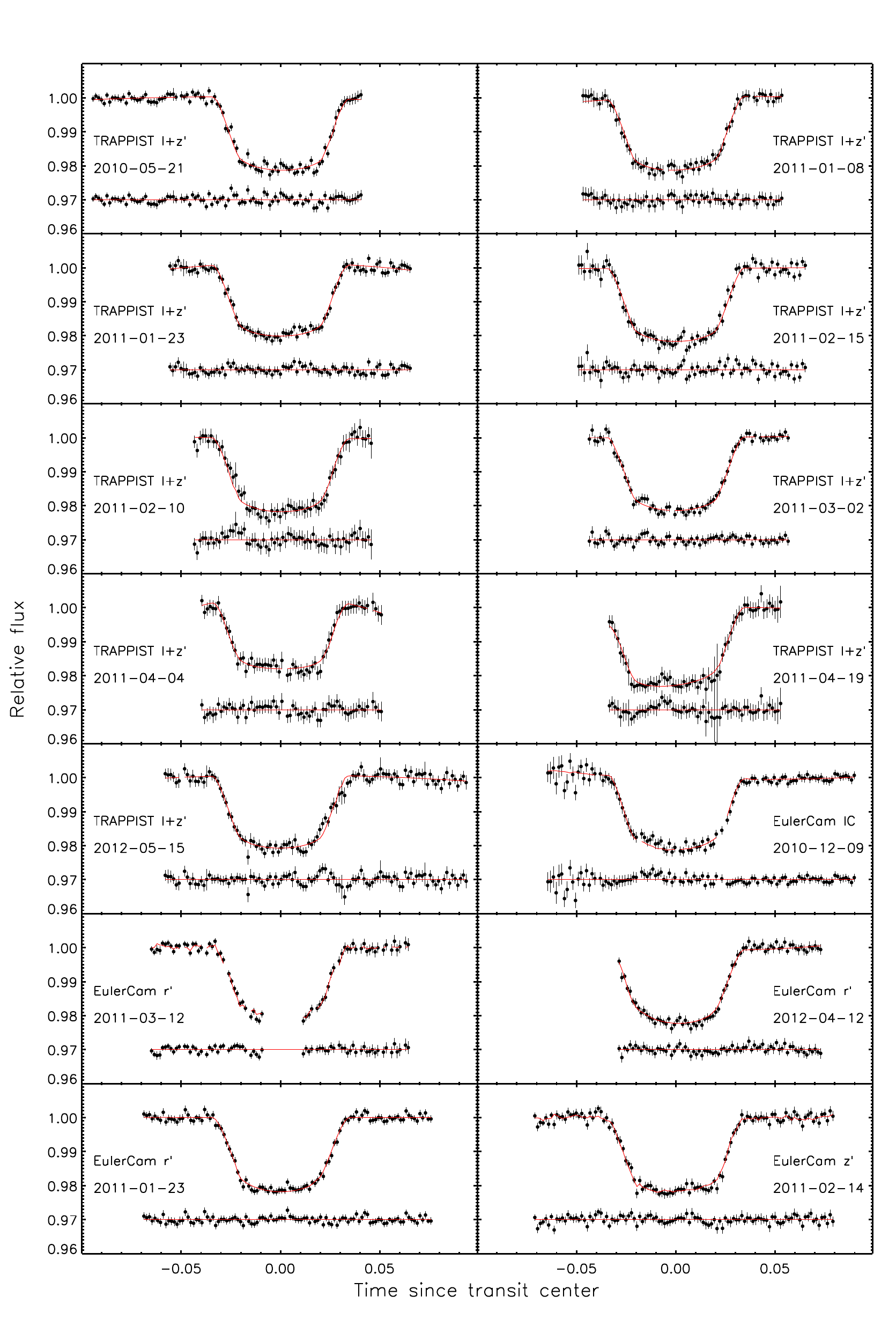}
\caption{\label{fig:lcs1}All transits observed together with their models and residuals. Instrument and filter used are indicated
in each frame. All data have been binned in two-minute intervals.}
\end{figure*}

\section{Observations and data reduction}
\label{sec:obs}

Between May 2010 and April 2012, we obtained a total of 25 lightcurves of WASP-19. Fourteen of these observations
were timed to observe the transit, while 11 were performed during the occultation of the planet. We made use of 
three instruments: EulerCam at the 1.2~m Euler-Swiss Telescope and the automated 0.6~m TRAPPIST Telescope at ESO La Silla 
Observatory (Chile), as well as HAWK-I at the VLT/UT4 at ESO Paranal Observatory (Chile). We include in our analysis also the Faulkes South Telescope (FTS)
lightcurve published by \citet{Hebb10}, the HAWK-I H-band observation by \citet{Anderson10b} and the radial velocity 
measurements presented in \citet{Hebb10} and \citet{Hellier11a}. All new observations are summarized in Table \ref{tab:phot}.

\begin{table*}
\centering                        
\begin{tabular}{c c c c c c c}        
\hline\hline                
  Date (UT) & Instrument & Filter & Eclipse Nature & Photometric Model Function & $\beta_{red}$& RMS [ppm, rel. flux, per 5 min] \T \\   
\hline 
  2010-01-20  & HAWK-I   & NB1190 & Occultation& $p(t^{2})$ and $p(t^{2})+p(FWHM^{1})$ & 1.23 and 1.11 & $1820$ and $1096$ \T  \\ 
  2010-05-21  & TRAPPIST & I+z' & Transit    & $p(t^{2})$ & 1.17 & $680$ \\  
  2010-12-09  & EulerCam & IC   & Transit    & $p(t^{2})+p(sky^{1})$ & 1.30 & $900$  \\  
  2011-01-08  & TRAPPIST & I+z' & Transit    & $p(t^{2})$ & 1.62 & $610$ \\  
  2011-01-21  & TRAPPIST & z'   & Occultation& $p(t^{2})$ & 1.00 & $720$\\  
  2011-01-23  & EulerCam & r'   & Transit    & $p(t^{2})$ & 1.22 & $570$\\ 
  2011-01-23  & TRAPPIST & I+z' & Transit    & $p(t^{2})$ & 1.42 & $730$ \\  
  2011-02-10  & TRAPPIST & I+z' & Transit    & $p(t^{2})$ & 2.05 & $1170$\\ 
  2011-02-14  & EulerCam & z'   & Transit    & $p(t^{2})+p(sky^{1})$ & 1.25 & $700$\\
  2011-02-15  & TRAPPIST & I+z' & Transit    & $p(t^{2})$ & 1.00 & $840$\\ 
  2011-02-23  & TRAPPIST & z'   & Occultation& $p(t^{2})$ &1.51  & $960$\\ 
  2011-02-24  & TRAPPIST & z'   & Occultation& $p(t^{2})$ & 1.00 & $690$\\ 
  2011-03-02  & TRAPPIST & I+z' & Transit    & $p(t^{2})$ & 1.00  & $610$\\ 
  2011-03-12  & EulerCam & r'   & Transit    & $p(t^{2})+p(FWHM^{2})$& 1.38 & $690$\\  
  2011-04-04  & TRAPPIST & I+z' & Transit    & $p(t^{2})$ & 1.82 & $980$ \\  
  2011-04-19  & TRAPPIST & I+z' & Transit    & $p(t^{2})$ & 1.80 & $1120$ \\ 
  2011-04-21  & TRAPPIST & z'   & Occultation& $p(t^{2})$ & 1.00 & $700$ \\   
  2011-04-28  & EulerCam & z'   & Occultation& $p(t^{2})+p(FWHM^{1})$ & 1.19  & $520$\\  
  2011-05-06  & EulerCam & z'   & Occultation& $p(t^{2})$ & 1.13 & $400$\\
  2012-02-28  & EulerCam & z'   & Occultation& $p(t^{2})$ & 1.00 & $510$\\ 
  2012-03-11  & EulerCam & z'   & Occultation& $p(t^{2})$ & 1.13 & $440$\\
  2012-03-15  & EulerCam & z'   & Occultation& $p(t^{2})+p(xy^{1})$ & 1.29  & $800$\\ 
  2012-03-18  & EulerCam & z'   & Occultation& $p(t^{2})+p(FWHM^{1})$ & 1.00 & $460$\\  
  2012-04-12  & EulerCam & r'   & Transit    & $p(t^{2})$ & 1.55 & $940$\\  
  2012-05-15  & TRAPPIST & I+z' & Transit    & $p(t^{2})$ & 1.24 & $970$\\  
\hline                                  
\end{tabular}
\caption{\label{tab:phot}A summary of newly obtained photometry. Date, instrument, filter, and the nature of eclipse are given for each observation together with 
the photometric model function, red noise amplitude $\beta_{red}$ \mbox{\citep[as defined in][]{Winn08}} and the RMS of the binned (5 minutes) residuals. The
notation $p(j^{i})$ refers to a polynomial of degree i of parameter j, e.g. $p(t^{2})$ denotes a polynomial of second degree with respect to time.}
\end{table*}

   \subsection{EulerCam}
   
Five transit and six occultation lightcurves have been obtained with EulerCam, the imager of the Euler-Swiss telescope at La Silla. The instrument and the
reduction of EulerCam data are described in detail by \citet{Lendl12}. The observations were done either with a focused telescope or applying a 
small $\le$ 0.1~mm defocus yielding stellar PSFs with a typical full width at half-maximum (FWHM) between 1.1 and 2.5 arcsec, while the exposure times were between 60s and 120s, 
depending on filter and conditions. On 12 March 2011, the CCD temperature during the observations was slightly elevated, $-100^\circ$~C instead of the nominal 
$-115^\circ$~C. The lightcurves were extracted using relative aperture photometry, with the reference stars and apertures selected independently for each observation.

   \subsection{TRAPPIST}

Nine transits and four occultations of WASP-19b were observed with the robotic 60~cm TRAPPIST \citep{Gillon11a,Jehin11} that is 
also located at the La Silla site. We defocused the telescope slightly in order to spread the light over more pixels yielding typical FWHM values between 3.6 and 6.5 arcsec on the images 
and used exposure times between 15s and 40s. Again, the lightcurves were obtained with relative aperture photometry, where IRAF \footnote[1]{IRAF is distributed by the National Optical Astronomy 
Observatories, which are operated by the Association of Universities for Research in Astronomy, Inc., under cooperative agreement with the National Science Foundation.} 
is used in the reduction process.

   \subsection{HAWK-I}

We obtained photometry of WASP-19 using the HAWK-I instrument \citep{Pirard04,Casali06} on VLT UT4 during two occultations of WASP-19b using the narrow band
filters NB1190 and NB2090. Unfortunately, during the NB2090 observations, the target exceeded the linearity range of the detector, as
a consequence the data do not have the necessary precision to detect the occultation; however, the NB1190 data are good, and thus we restrict our analysis to them.

The NB1190 observations of WASP-19 took place on 20 January 2010 from 02:55 to 06:55 UT covering the predicted occultation time together 
with 145 min of observations outside of eclipse. The detector integration time (DIT) was kept short (3s) so the counts of the target and a bright reference star
did not exceed the linear range of the detector. The data were obtained by alternating between two jitter positions, in order to be able to 
correct for background variations if necessary. 

\begin{figure*}
\includegraphics{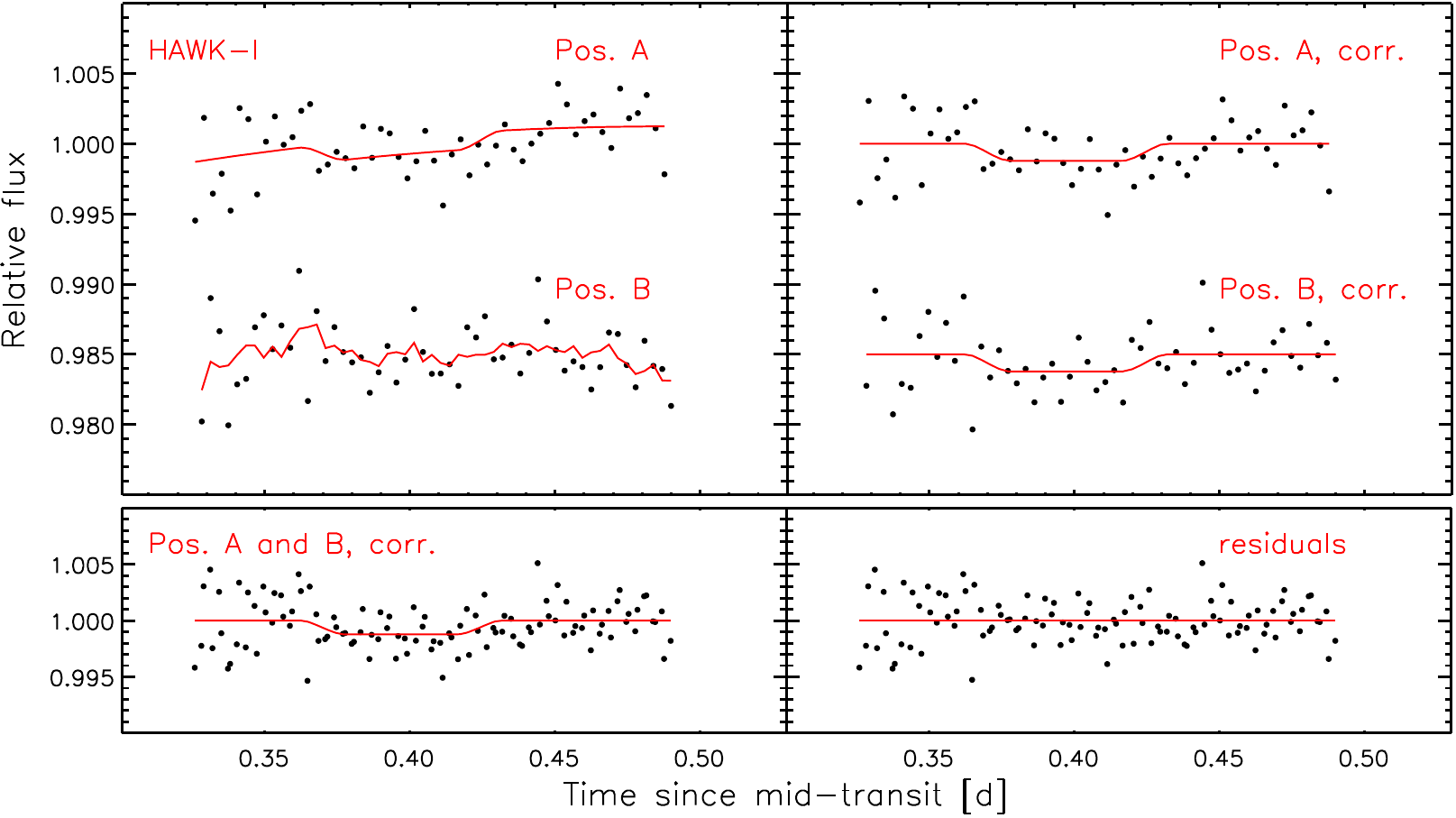}
\caption{\label{fig:lcs3}The occultation observed with HAWK-I at 1.19 {\mym} on 20 January 2010. The lightcurves obtained
from the two jitter positions are shown. Top panel: the raw lightcurves, together with the occultation and photometric model (left),
and divided by the photometric model (right). Bottom panel: both lightcurves corrected and normalized with 
the best fit (left) and the residuals (right).}
\end{figure*}

The data were corrected for dark and flat field effects using standard procedures. Then, we identified bad pixels on the images and 
substituted their values by the mean of the neighboring pixels. Here, we experimented with different cutoffs
for the identification of bad pixels and obtained the best results discarding pixels deviating by 4~$\sigma$ for background values and 40~$\sigma$
for stars. The target flux was extracted from the corrected images using aperture photometry. We tested a set of constant apertures, as well
as apertures that varied from image to image as a function of the stellar FWHM. The sky annulus was kept constant for all images.
The best result was obtained using a variable aperture of two times the FWHM. We tested all bright stars of the four HAWK-I chips 
and found the best photometry using only the single bright star located on the same chip as WASP-19. The bright stars on the other
detectors showed significantly different variations so were not used. The lightcurves are shown in Figure \ref{fig:lcs3}.

\begin{figure*}
\includegraphics{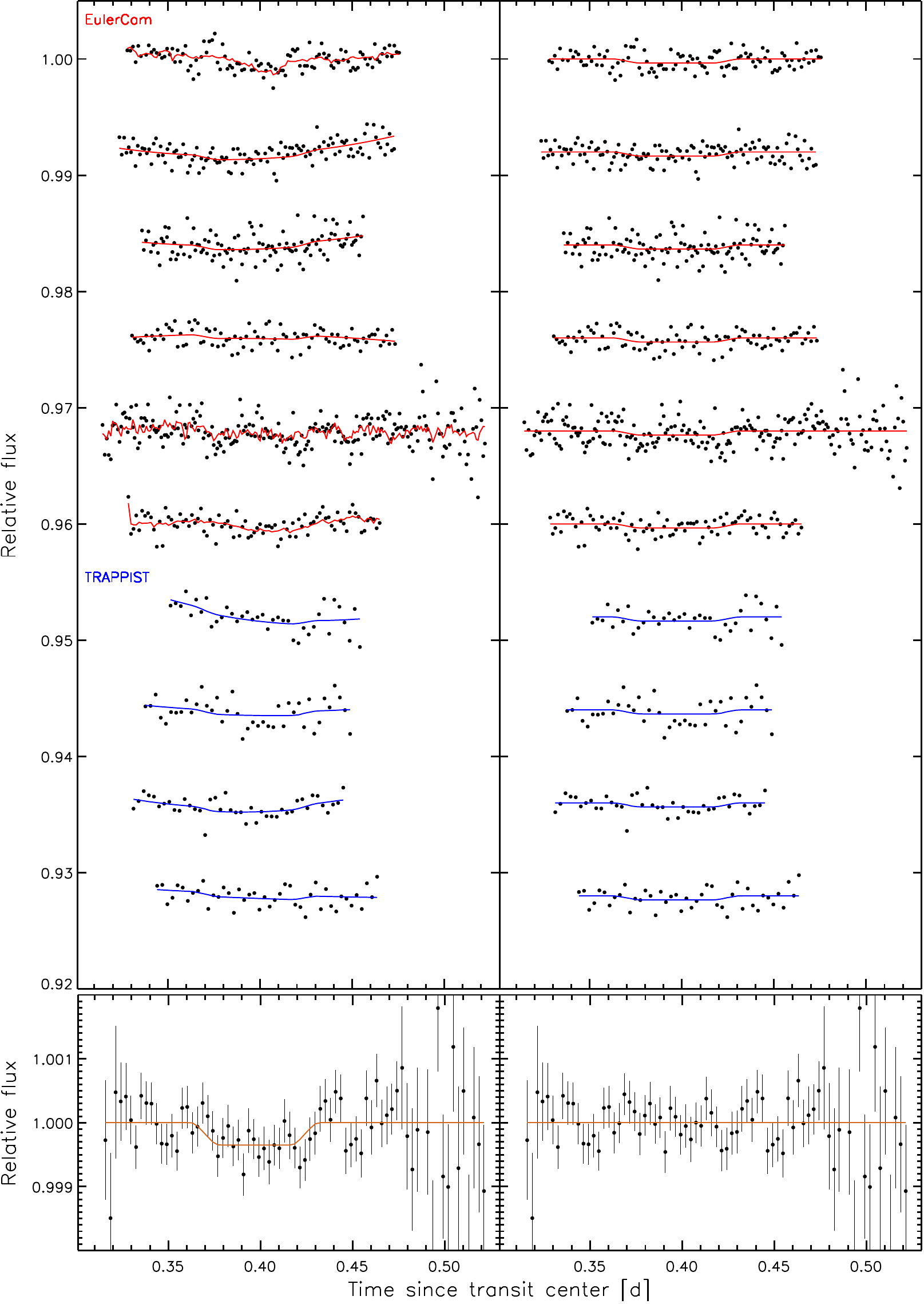}
\caption{\label{fig:lcs2}All occultation lightcurves obtained in the z'-band. The upper six lightcurves were obtained with EulerCam
and are unbinned, while the lower four lightcurves were obtained with TRAPPIST and binned in two-minute intervals. Left panel:
the raw lightcurves, together with the occultation and photometric models. Right panel: the lightcurves and the 
occultation model, divided by the photometric model. The lower panel shows all data, corrected for the photometric 
model and binned in four-minute intervals. Data and model are shown on the left, while the residuals are shown on the right.}
\end{figure*}

\section{Modeling}
\label{sec:mod}
We performed a combined analysis of all photometric (transit and occultation) data, together with published radial velocities.
The data were modeled using the Markov chain Monte Carlo (MCMC) method in order to derive the posterior probability distributions
of the parameters of interest (see \ref{sec:mcmc} for details). Incorporated in our analysis are models for photometric correction
functions, which account for photometric variations not related to the eclipse, i.e. airmass, weather, or instrumental
effects. We also rescale our error bars if they show to be underestimated. Please see Section \ref{sec:bas} for details.

\subsection{MCMC}
\label{sec:mcmc}
We employed the MCMC method using the implementation
described in \citet{Gillon10a,Gillon12a}. In short, the radial velocities are modeled using a Keplerian orbit and with the 
prescription of the Rossiter-McLaughlin effect \citep{Rossiter24,McLaughlin24} provided by \citet{Gimenez06}. The 
photometric model for eclipses (transits and occultations) is that of \citet{Mandel02}, used without limb darkening for occultations. 
The jump parameters are transit depth $dF$, impact parameter $b$, 
transit duration $d$, time of midtransit $T_{0}$, period $P$, occultation depths $dF_{occ}$ for each wavelength,
and $K_{2} = K\sqrt{1-e^{2}}P^{1/3}$ (where $K$ and $e$ denote 
the radial velocity semi-amplitude and eccentricity, respectively). The jump parameters $\sqrt{e}\cos\omega$ and $\sqrt{e}\sin\omega$ 
(where $\omega$ denotes the argument of periastron) are used to determine of the eccentricity. Limb darkening
is accounted for by using the combinations $c_{1} = 2u_{1} + u_{2}$ and $c_{2} = u_{1} - 2u_{2}$ of the calculated 
limb-darkening coefficients of \citet{Claret11} following \citet{Holman06}. With the exception of the limb darkening
parameters (for which we use a normal prior with a width equal to the error quoted by \citealp{Claret11}), we assume uniform prior 
distributions. We followed the method described
by \citet{Enoch10} using the mean stellar density, temperature, and metallicity for determining the stellar mass and radius.
In our whole analysis, we always ran at least two MCMC chains and checked convergence with the Gelman \& Rubin test \citep{Gelman92}.
All time stamps are converted to the TDB time standard, as described by \citet{Eastman10}.

\subsection{Photometric model and error adaptation}
\label{sec:bas}
As described in Section \ref{sec:int}, ground-based lightcurves are often affected by red noise correlated with external
parameters. In our MCMC analysis, we have the possibility to include time, FWHM, coordinate shifts, and
background variations in our model. This is done by multiplying the transit model by a polynomial (up to 4th degree) with respect to any combination
of these parameters. The coefficients of the polynomial are not included as jump parameters in the MCMC but are found by minimization of the 
residuals at each step. In order to account for airmass and stellar variability effects, we assumed a second-order polynomial with 
respect to time as the minimal accurate model for ground-based photometry. 
We checked more complex models by running MCMC chains of $10^{5}$ points on each lightcurve including higher orders of time dependence and additional 
terms in FWHM, pixel position and background. A more complex model was favored over a simple one only if the 
Bayes factor \citep{Schwarz78} estimated from the Bayesian information criterion indicated a significantly higher probability (i.e. $B_{1,2} > 100$). The best photometric 
model functions are listed in Table \ref{tab:phot} and were used in all subsequent analyses.
For the H-band data, we kept the coordinate dependence described in \citet{Anderson10b}, and the archive FTS lightcurve 
\citep{Hebb10} was fitted with the minimal model.
 
Although photometric error bars are usually derived including scintillation, readout, background, and photon noise, they 
are often underestimated, and we adapted them by accounting for additional white and red noise. For the white noise, we derived 
a scaling factor $\beta_{w}$ from the ratio of the mean photometric error and the standard derivation of the photometric residuals.
For the red noise, we obtained a scaling factor $\beta_{r}$ by comparing the standard deviation of the binned photometric residuals
to the standard deviation of the complete dataset, as described in detail by \citet{Winn08} and \citet{Gillon10a}. 
Finally we multiplied both scaling factors to obtain the correction factors $CF = \beta_{w} \times \beta_{r}$ for the photometric errors.
In the subsequent analysis, all photometric error bars were multiplied by these factors.
Analogously, we computed values for the radial velocity jitter that were added quadratically to the radial velocity errors.

\subsection{Summary of tested models}
Having derived the above factors, we analyzed the entire 
dataset. We did so by running chains of $10^{5}$ points on all photometric and radial velocity data. Next to the global analysis, we also performed analyses
of subsets of lightcurves (described in detail in Sections \ref{sec:res:tra} and \ref{sec:res:occ}).
Additionally, we searched for any color dependence in the transit depths by allowing for depth offsets between the 
different filters (Section \ref{sec:res:tdv}) and also derived individual midtransit times in order to check for transit-timing 
variations (Section \ref{sec:res:tdv}). To verify the result of \citet{Burton12}, we performed a global analysis in which we 
included their value, $DF_{occ,z'} = 880 \pm 190$~ppm, as a Gaussian prior (Section \ref{sec:res:occ}).
The results are described in detail in Section \ref{sec:res}, while all newly obtained lightcurves are shown in Figure \ref{fig:lcs1} 
(transits), Figure \ref{fig:lcs3} (NB1190 occultation), and Figure \ref{fig:lcs2} (z'-band occultations).

\section{Results}
\label{sec:res}

\subsection{The TRAPPIST transit sequence}
\label{sec:res:tra}

\label{sec:TRA}
\begin{figure*}
\includegraphics{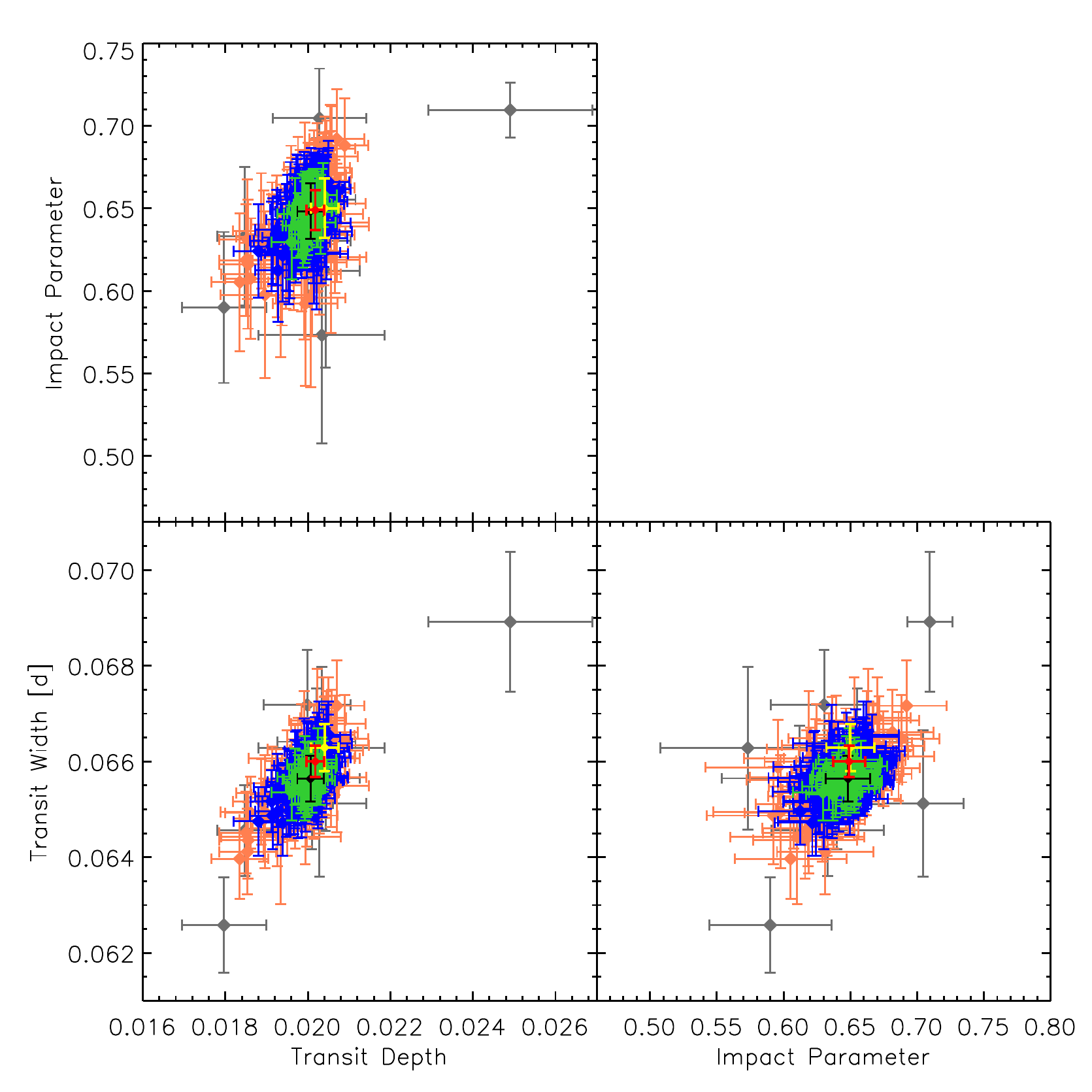}
\caption{\label{fig:TRA}The solutions found from fits to all possible subsets of TRAPPIST transits are shown. The single transits are shown in gray,
and combinations of three, five and seven transits are shown in orange, blue and green, respectively. The solution to all TRAPPIST and EulerCam lightcurves 
are shown in black and yellow, respectively. The global solution is shown in red.}
\end{figure*}

To investigate the benefits of combining several lightcurves, we divided our set of nine I+z' TRAPPIST transit lightcurves 
into subsets containing all possible combinations of one to nine lightcurves and performed an MCMC analysis on each of them using the procedure described 
in Section \ref{sec:mod}. We can observe how the solutions converge as we use an increasing number of transits 
in Figure \ref{fig:TRA}. The outlier located at high transit depth stems from the transit observed on 10 February 2011, which
is showing very high red noise. It is clearly visible that combinations favoring a large transit depth also
require a larger transit width. This correlation might be related to the presence of star spots occulted by the planet.
While spots on the limb of the star shorten the apparent transit duration, spots closer to the center of the star will produce a flux increase leading to a decrease
in the apparent transit depth. It is also possible that the overall stellar variability is not well constrained, e.g. due to little or no out-of-transit data, and thus
the photometric correction model cannot be determined correctly. Figure \ref{fig:TRA} also shows the global solution and the solution obtained from the EulerCam
transits alone. The results from TRAPPIST and EulerCam agree within their error bars, yet the EulerCam data find a slightly larger transit
depth, width, and impact parameter. 

\begin{figure}
\includegraphics[width=0.95\linewidth]{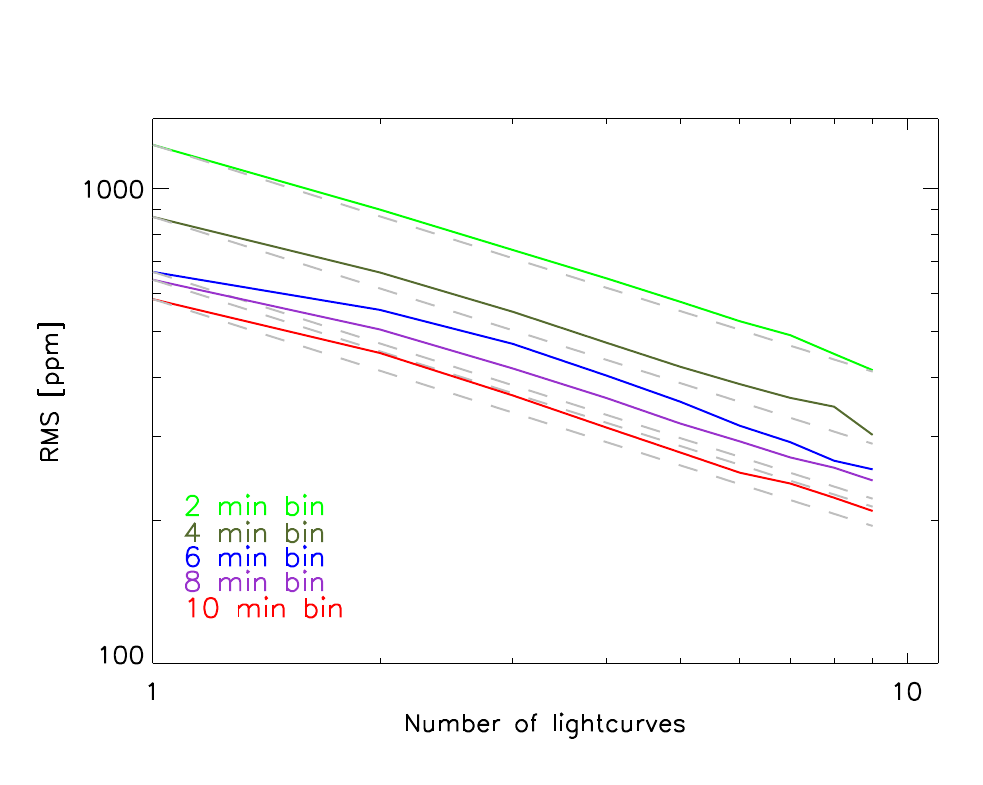}
\caption{\label{fig:STD}The increase in photometric precision by combining up to nine lightcurves from TRAPPIST. The RMS in bin sizes of (from top to bottom)
two, four, six, eight and ten minutes are shown. The gray dashed lines show the expected $1/\sqrt{n_{lc}}$ decrease for each time bin.}
\end{figure}

Next to investigating the parameters obtained from combinations of transits, one can also check the photometric improvement reached by combining an increasing number 
of lightcurves. To measure this effect, we used each set of combinations of $n=1$ to $n=9$ lightcurves, folded them on the best-fit period, and 
binned the data. Points before -0.05 and after +0.05 days from transit center were discarded to avoid phases that are not covered by all lightcurves.
In Figure \ref{fig:STD}, we show the photometric RMS against the number of combined lightcurves and compare the decrease in RMS to the ``best case''
(decrease with $\frac{1}{\sqrt{n}}$). The increase in precision is near that value, particularly if small time bins are used. By combining all 
nine TRAPPIST lightcurves, we obtain an RMS of 321 ppm for a moderately sized time bin of five minutes.

\subsection{One simultaneous observation}

\begin{figure}[h!]
\includegraphics[width=\linewidth]{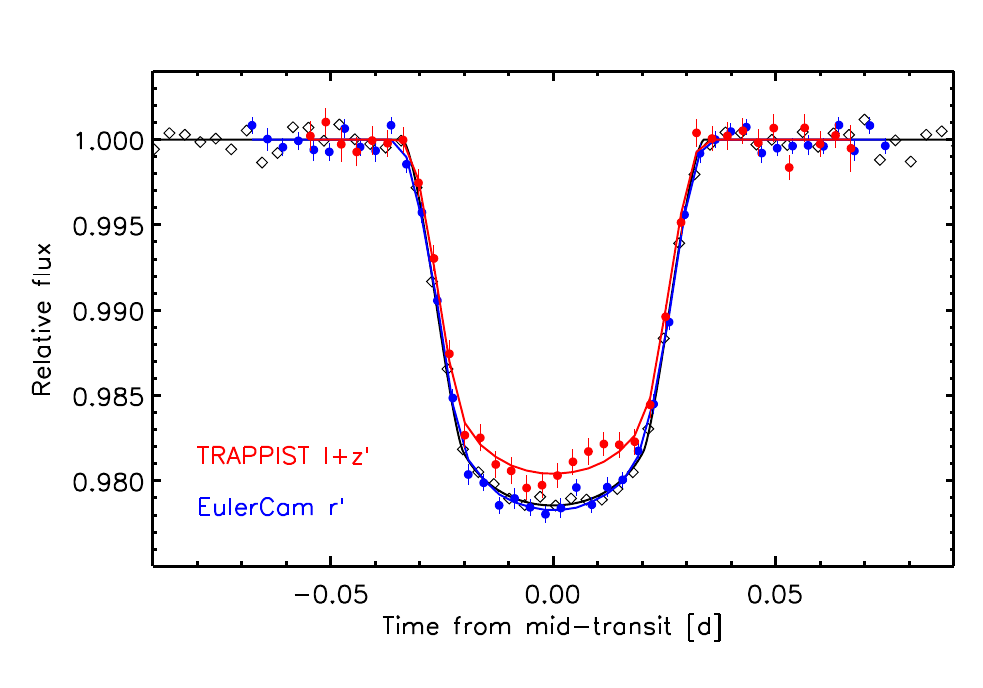}
\caption{\label{fig:TTR} The simultaneous transit observation performed with TRAPPIST (red) and EulerCam (blue). For clarity, the data are binned into 
5-minute bins, and the models of the individual transits are shown as solid lines. The binned data from the combination of all I+z' lightcurves 
(same as in Figure \ref{fig:trac}) are shown in black for comparison.}
\end{figure}

The transit of 23 January 2011 was observed with EulerCam and TRAPPIST simultaneously, using an \textit{r'-Gunn} filter on EulerCam and 
an \textit{I+z'} filter on TRAPPIST.
Figure \ref{fig:TTR} depicts the two lightcurves superimposed. It is obvious that the TRAPPIST data 
are showing an anomalously small transit depth and a short-term brightening during the second half of the transit.
With only one lightcurve, one might conclude that the planet crossed a star spot during transit. Since the EulerCam light 
curve was observed at shorter wavelengths and given the cooler temperature of the spot, we would expect the effect to be 
more pronounced here. However, the feature is absent in the r'-band, excluding the spot-hypothesis. We tried to
account for it by adopting photometric correction models including backgound, sky, and FWHM parameters, yet were
not able to find a model that reproduced the lightcurve shape.
Removing the points during the second half of the transit in the TRAPPIST lightcurve, we obtain a value within 1 $\sigma$ of the EulerCam result.

\subsection{Global Analysis}

We performed a global MCMC analysis using all available radial velocity and photometric data including transits and occultations
(described in detail in Section \ref{sec:mod}). The results are shown in Table \ref{tab:res}. 

\begin{table*}
\centering                        
\begin{tabular}{p{6cm} p{5cm} p{5cm}}       
\hline\hline 
  \multicolumn{3}{c}{WASP-19} \T \\    
\hline   
 \multicolumn{3}{l}{Jump parameters} \T  \\
\hline
Transit depth &  $\Delta F = (R_{p}/R_{\ast})^{2}$ & $0.02018\pm0.00021$ \T  \\
 & $ b' = a*\cos(i_{p})$ $[R_{\ast}]$ & $0.649\pm0.012$  \\
Transit duration & $ T_{14}$ [d] & $0.06586_{-0.00031}^{+0.00033}$  \\
Time of midtransit & $ T_{[0]} - 2450000$ [HJD] & $6029.59204\pm0.00013$\\
Period & $ P$ [d] & $0.7888390\pm 2\times 10^{-7}$ \\
 & $ K_{2} = K\sqrt{1-e^{2}}P^{1/3} $ [{\msd}] & $238.1\pm2.7$  \\
z'-band occultation depth & $ \Delta F_{occ z'}$ [ppm] & $352\pm116$ \\
NB1190 occultation depth & $ \Delta F_{occ NB1190}$ [ppm] & $1711_{-726}^{+745} $ \\
H-band occultation depth & $ \Delta F_{occ H}$ [ppm] & $3216_{-455}^{+473} $ \\
 & $ \sqrt{e}$ cos $\omega$ & $0.053\pm0.020$ \\
 & $ \sqrt{e}$ sin $\omega$ & $0.054_{-0.082}^{+0.057} $ \\
 & $ \sqrt{v_{\ast}\sin{I_{\ast}}}$ cos $\beta$ & $1.85_{-0.19}^{+0.17}$\\
 & $ \sqrt{v_{\ast}\sin{I_{\ast}}}$ sin $\beta$ & $-0.27\pm0.23$\\
 & $ c_{1,\rm r'} = 2u_{1,r} + u_{2,r} $ & $1.123\pm0.040$ \\
 & $ c_{2,\rm r'} = u_{1,r} - 2u_{2,r} $ & $-0.052\pm0.037$ \\
 & $ c_{1,\rm IC} = 2u_{1,IC} + u_{2,IC} $ & $ 0.903\pm0.034$ \\
 & $ c_{2,\rm IC} = u_{1,IC} - 2u_{2,IC} $ & $-0.173\pm0.025$ \\
 & $ c_{1,\rm I+z'} = 2u_{1,I+z'} + u_{2,I+z'} $ &  $0.840\pm0.050$ \\
 & $ c_{2,\rm I+z'} = u_{1,I+z'} - 2u_{2,I+z'} $ & $-0.251\pm0.053$ \\
 & $ c_{1,\rm z'} = 2u_{1,z'} + u_{2,z'} $ & $0.831\pm0.029$ \\
 & $ c_{2,\rm z'} = u_{1,z'} - 2u_{2,z'} $ & $-0.218\pm0.021$ \\
\hline 
 \multicolumn{3}{l}{Deduced parameters} \T \\
\hline
 Radial velocity semi-amplitude & $ K $ [{\ms}] \T & $257.7\pm2.9$ \\
 Planet radius & $ R_{p} $ [{\Rjup}] & $1.376\pm0.046$  \\
 Planet mass & $ M_{p} $ [{\Mjup}] & $1.165\pm0.068$ \\
 Planet density & $ \rho_{p} $ [{\rhojup}]  & $0.447_{-0.025}^{+0.027}$ \\
 Eccentricity & $ e $ & $0.0077_{-0.0032}^{+0.0068}$ \\
 Argument of periastron & $ \omega $ [deg] & $43_{-67}^{+28}$ \\
 Semi-major axis & $ a $ [AU] & $ 0.01653 \pm0.00046$ \\
 Normalized semi-major axis & $ a/R_{\ast} $ & $3.573\pm0.046$ \\
 Inclination & $ i_{p} $ [deg] & $79.54\pm0.33$ \\
 Transit impact parameter & $ b_{tr} $ & $0.645\pm0.012$  \\
 Occultation impact parameter & $ b_{occ} $ & $0.652\pm0.015$  \\
 Time of midoccultation & $ T_{occ} - 2450000$ [HJD] & $6030.77766\pm0.00088$  \\
 Projected spin-orbit angle & $ \beta$ [deg] & $-8.4_{-7.2}^{+7.0}$ \\ 
 Planet equilibrium temperature\tablefootmark{a} & $ T_{eq} $ [K] & $2058\pm40$  \\
 Planet surface gravity & $ \log g_{p} $ [cgs] & $3.184\pm0.015$ \\ 
 Stellar mass & $ M_{\ast} $ [{\Msolar}]  & $0.968_{-0.079}^{+0.084}$ \\
 Stellar radius & $ R_{\ast} $ [{\Rsolar}]  & $0.994\pm0.031$ \\
 Stellar mean density & $ \rho_{\ast} $ [{\rhosun}] & $0.983_{-0.036}^{+0.039}$  \\
 1\textsuperscript{st} quadratic LD coeff., r' band & $ u_{1,\rm r'} $ & $0.439\pm0.020$\\
 2\textsuperscript{nd} quadratic LD coeff., r' band & $ u_{2,\rm r'} $ & $0.246\pm0.015$\\
 1\textsuperscript{st} quadratic LD coeff, I+z' band & $ u_{1,\rm I+z} $ & $0.286\pm0.026$ \\
 2\textsuperscript{nd} quadratic LD coeff., I+z' band & $ u_{2,\rm I+z} $ & $0.268\pm0.019$ \\
 1\textsuperscript{st} quadratic LD coeff., IC band & $ u_{1,\rm IC} $ & $0.326\pm0.016$ \\
 2\textsuperscript{nd} quadratic LD coeff., IC band & $ u_{2,\rm IC} $ & $0.2497\pm0.0094$ \\
 1\textsuperscript{st} quadratic LD coeff., z' band & $ u_{1,\rm z'} $ & $0.289\pm0.014$ \\
 2\textsuperscript{nd} quadratic LD coeff., z' band & $ u_{2,\rm z'} $ & $0.2536\pm0.0077$ \\ 
\hline
\end{tabular}
\caption{\label{tab:res} The median values and the 1-$\sigma$ errors of the marginalized posterior PDF 
obtained from the MCMC analysis of all new data plus the radial velocities and lightcurves published 
by \citet{Hebb10}, \citet{Anderson10b} and \citet{Hellier11a}.
\newline \tablefoottext{a}{Assuming an Albedo of A=0 and full redistribution from the planets day to night side, F=1.}}
                            
\end{table*}

\subsubsection{Eccentricity}

While \citet{Hebb10} and \citet{Hellier11a} did not measure a significant nonzero eccentricity from the analysis of 
radial velocity and transit data, 
\citet{Anderson10b} presented for WASP-19b a value of $ e= 0.016^{+0.015}_{-0.007}$ from the timing of their H-band occultation.
When including their data in our analysis, we derived a lower value for the eccentricity with a similar significance $ 0.0077_{-0.0032}^{+0.0068}$.
Removing the H-band data, we obtained $0.0061_{-0.0043}^{+0.0063}$, clearly not significant. We therefore do not see any evidence for a 
nonzero eccentricity of the WASP-19 system. 

\subsubsection{Transit depth and timing variations}
\label{sec:res:tdv}

\begin{table}
\centering                        
\begin{tabular}{ccc}       
\hline\hline 
 Filter & Wavelength (width) [nm]&$R_{p} / R_{\ast}$ \T  \\
\hline
 r'-Gunn & 664 (99) & $0.1437\pm0.0013$ \T  \\
 IC & 760 (91) & $0.1336\pm 0.0026$\\
 I+z' & 838 (190) & $0.14171\pm 0.00094$ \\
 z'-Gunn & 912.4 (68) & $0.1428\pm0.0014$ \\
\hline
\end{tabular}
\caption{\label{tab:ddf}The planet/star radius ratios obtained from a global MCMC analysis, while allowing for filter-dependent transit depths. For the I+z' and z' filters, 
the widths given here have been determined from the combination of the filter transmission and detector response curves. For the r' and IC filters we give the 
equivalent widths.}  
\end{table}

\begin{table}
\centering                        
\begin{tabular}{r c c c}       
\hline\hline 
Epoch & Mid-transit time & O-C [min] & deviation in $\sigma$ \T \\
      &[$HJD_{TDB}-2450000$] & & \\
\hline
$-1537 $&$4817.14633\pm0.00021$ &$-0.24\pm0.30$ & $0.8$ \T\\
$ -876 $&$5338.56927\pm0.00023$ &$ 0.27\pm0.33$ & $0.8$ \\
$ -621 $&$5539.72327\pm0.00030$ &$ 0.36\pm0.44$ & $0.8$ \\
$ -583 $&$5569.69826\pm0.00036$ &$-0.92\pm0.53$ & $1.8$ \\
$ -564$\tablefootmark{a}&$5584.68693\pm0.00024$ &$ 0.12\pm0.34$ & $0.4$ \\
$ -564$\tablefootmark{b}&$5584.68684\pm0.00019$ &$-0.00\pm0.28$ & $0.0$ \\
$ -541 $&$5602.83138\pm0.00046$ &$ 1.78\pm0.67$ & $2.7$ \\
$ -536 $&$5606.77464\pm0.00022$ &$ 0.44\pm0.32$ & $1.4$ \\
$ -535 $&$5607.56241\pm0.00033$ &$-1.10\pm0.47$ & $2.3$ \\
$ -516 $&$5622.55057\pm0.00026$ &$-0.79\pm0.38$ & $2.1$ \\
$ -503 $&$5632.80612\pm0.00025$ &$ 0.14\pm0.36$ & $0.4$ \\
$ -474 $&$5655.68222\pm0.00045$ &$-0.19\pm0.66$ & $0.3$ \\
$ -455 $&$5670.66976\pm0.00064$ &$-0.78\pm0.93$ & $0.8$ \\
$    0 $&$6029.59250\pm0.00035$ &$ 0.67\pm0.51$ & $1.3$ \\
$   43 $&$6063.51174\pm0.00030$ &$-0.54\pm0.44$ & $1.2$ \\
\hline
\end{tabular}
\caption{\label{tab:ttv} The midtransit times and their deviation from a linear ephemeris. The values were 
obtained from the combined analysis of all transits while the ephemeris was fixed to the one quoted in Table \ref{tab:res}. 
\newline \tablefoottext{a}{TRAPPIST lightcurve}
\newline \tablefoottext{b}{EulerCam lightcurve}
}
\end{table}

\begin{figure}
\includegraphics[width=0.9\linewidth]{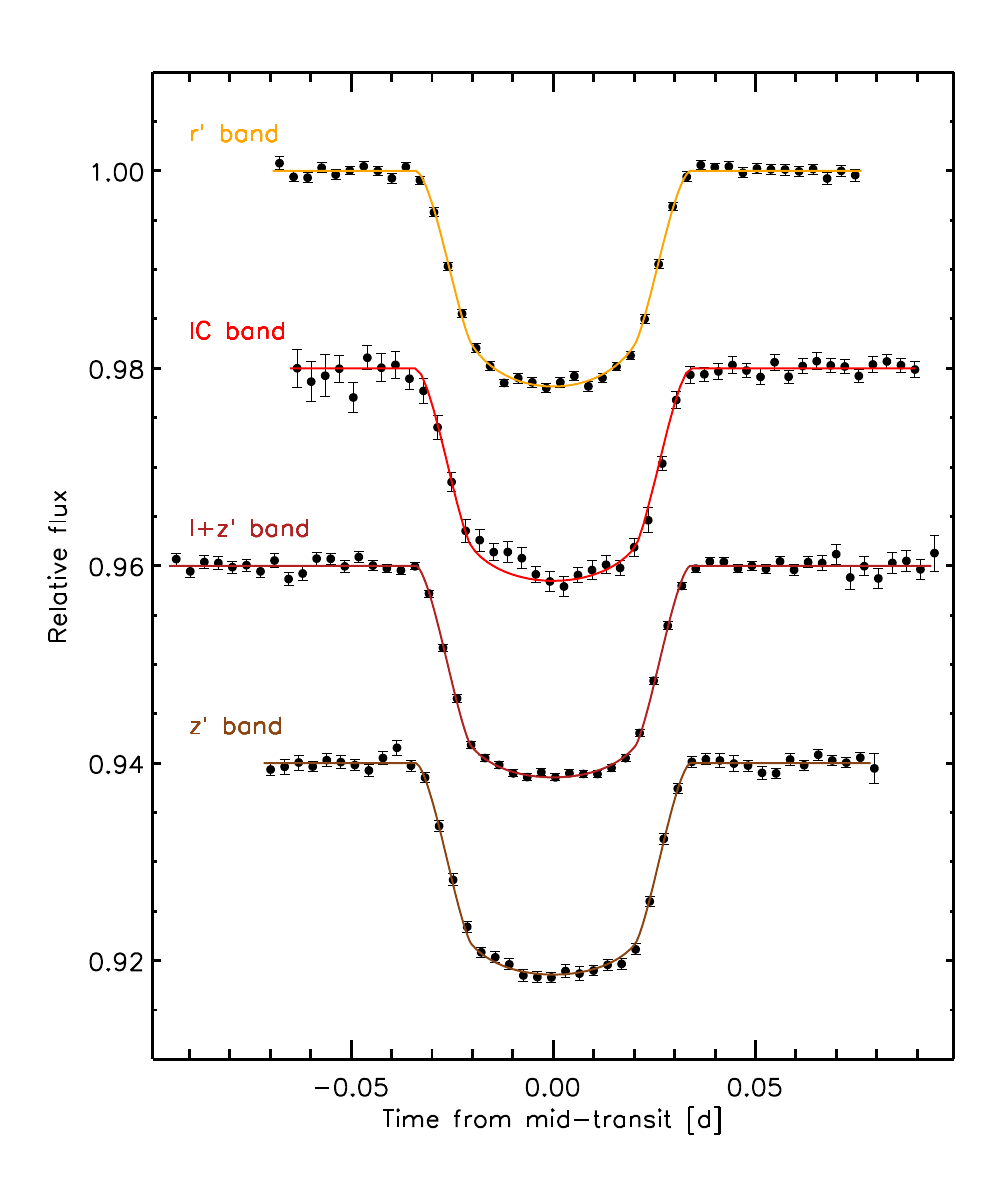}
\caption{\label{fig:trac}The combined lightcurves in each of the observed bands binned in five-minute intervals. The lightcurves
are a combination of (from top to bottom) three, one, nine, and two observations. Using the displayed five-minute intervals, the 
RMS of the residuals in the interval [-0.05,0.05] days are (from top to bottom) 432, 1021, 321, and 487 ppm.}
\end{figure}

\begin{figure}
\includegraphics[width=0.9\linewidth]{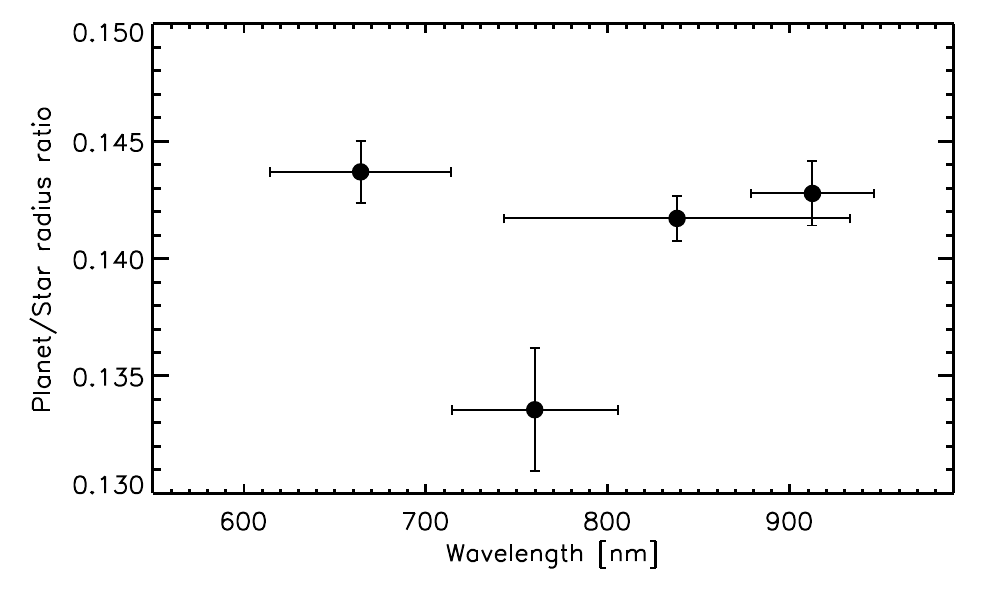}
\caption{\label{fig:ddf}The planet/star radius ratios obtained from an MCMC analysis of all data, while allowing for 
filter-dependent transit depths. The errorbars in wavelength span the filters effective width. The 
deviation IC lightcurve is based on only a single observation.}
\end{figure}

One of the checks we performed on the data was letting the transit depth vary for different filters, in order to 
search for any wavelength dependencies in the star/planet radii ratio which can be used to constrain models of the atmospheric 
transmission. The phase-folded and binned lightcurves for each filter are shown in Figure \ref{fig:trac}, and the respective radii ratios
are shown in Figure \ref{fig:ddf} and listed in Table \ref{tab:ddf}. The r', I+z', and z' band values match very well, only the IC value is 
slightly lower, 2.6 $\sigma$ below the global solution. 

\begin{figure}
\includegraphics[width=0.9\linewidth]{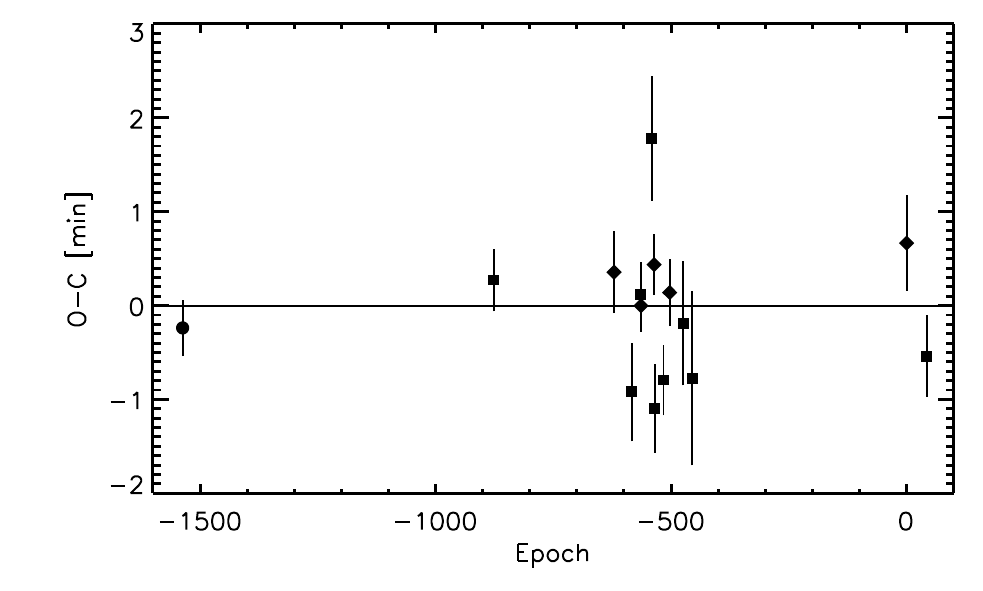}
\caption{\label{fig:ttv}The O-C deviations of the individual transits from the ephemeris 
given in Table \ref{tab:res} are shown. The filled circle represents the FTS transit of \citet{Hebb10}, the squares
represent data obtained with EulerCam, and the diamonds represent data from TRAPPIST.}
\end{figure}

We also searched for any deviation from a linear ephemeris by performing a global analysis while fixing the ephemeris to the 
ephemeris derived from the global analysis but letting the individual midtransit times vary. The results are shown in 
Figure \ref{fig:ttv} and listed in Table \ref{tab:ttv}. While there is some (expected) scatter around the linear ephemeris,
none of the deviations exceed 2.7~$\sigma$, so we find no evidence of TTVs in the WASP-19 system.

\subsubsection{z'-~band and 1.19 {\mym} occultations}
\label{sec:res:occ}

\begin{figure}[h!]
\includegraphics[width=0.9\linewidth]{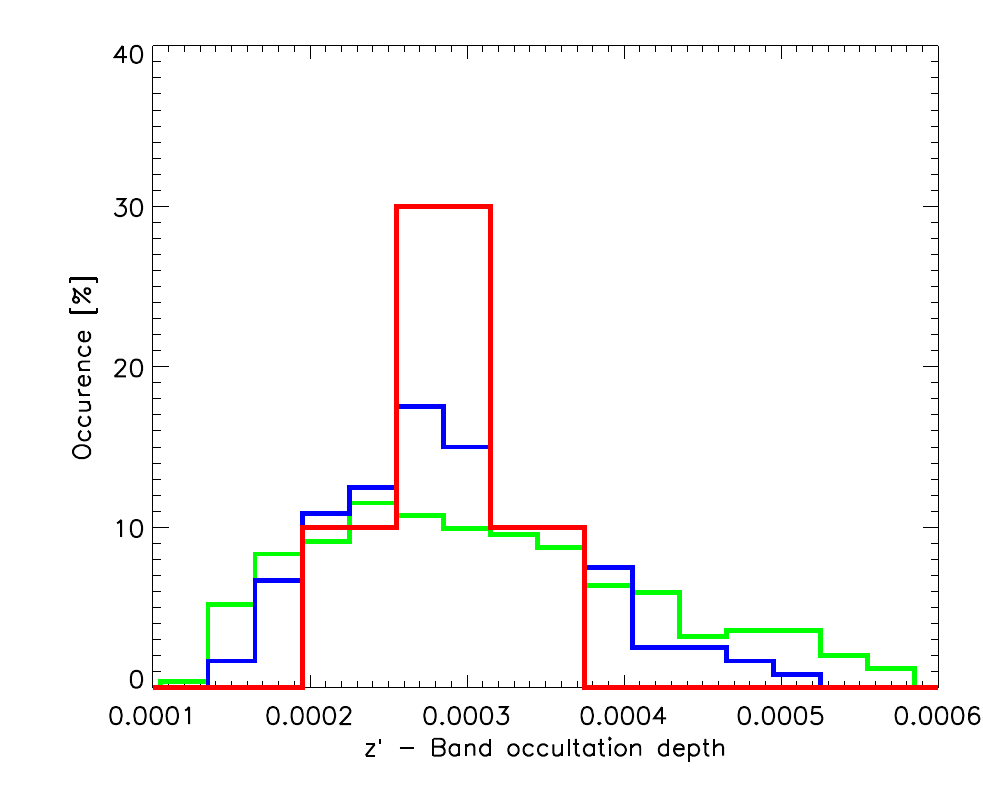}
\caption{\label{fig:his}A histogram of the results obtained from modeling combinations of five (green), seven (blue), and nine (red) occultations.}
\end{figure}

\begin{figure}[h!]
\includegraphics[width=0.9\linewidth]{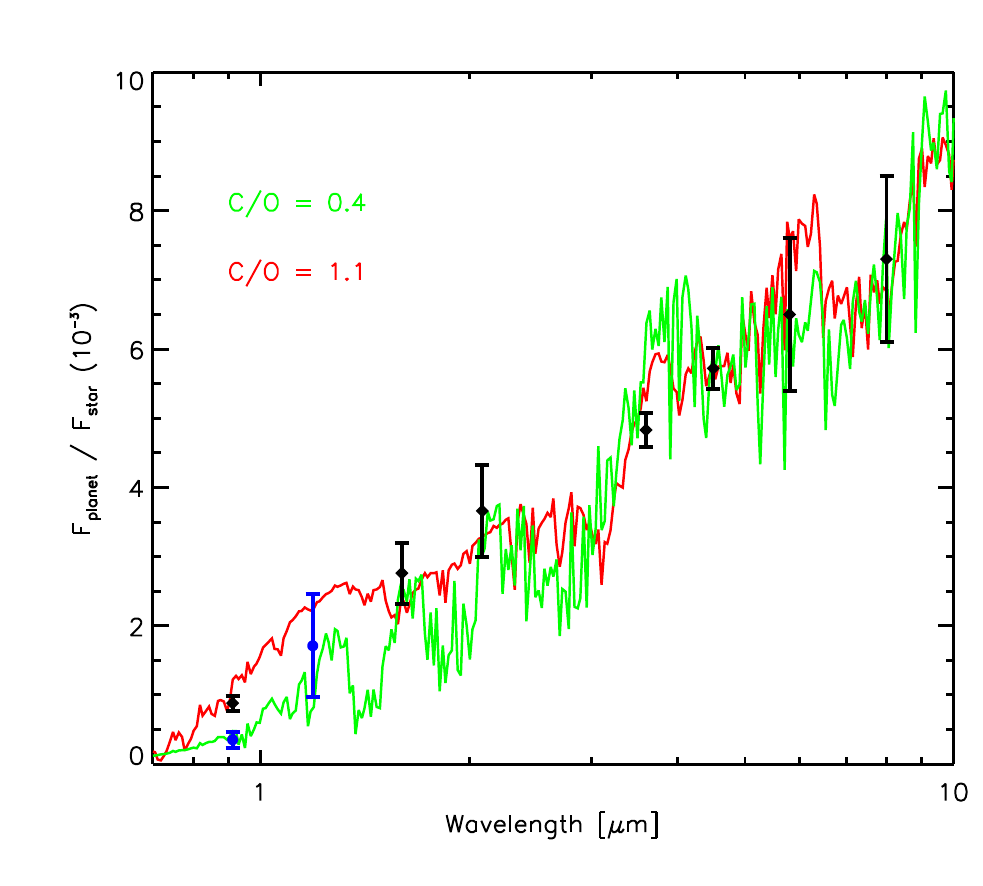}
\caption{\label{fig:atm} The model spectra of the dayside atmosphere of WASP-19b computed by \citet{Madhu12a} compared to observations. 
The z'-band and 1.19 {\mym} observations presented in this work are shown in blue, while the data of \citet{Anderson10b,Anderson11b,Burton12} 
and \citet{Gibson10} are shown in black. The two model atmospheres shown have been computed for the carbon-dominated (C/O = 1.1, red) and 
oxygen-dominated (C/O = 0.4, green) case and are reproduced here with kind permission of N. Madhusudhan.}
\end{figure}

We measure a z'-~band occultation depth of $352\pm116$~ppm from the combined analysis of the ten z'-~band occultation lightcurves 
in our dataset. For individual lightcurves the occultation is well buried in the noise. To verify that our nonzero
occultation depth is not caused by a systematic effect present in a single or a small number of lightcurves, we proceeded in a 
similar way to what we did with the set of TRAPPIST transits (Section \ref{sec:TRA}). We created all possible subsets containing at least 
five occultation lightcurves and analyzed them while fixing all parameters except the occultation depth to the values derived above. 
Histograms of the derived occultation depths are shown in Figure \ref{fig:his}. The results obtained from fits of fewer light 
curves are consistent with the presented value, although they have lower significance. Recently, \citet{Burton12} have presented an 
occultation depth for WASP-19b of $880\pm190$~ppm based on one lightcurve obtained with ULTRACAM mounted at the NTT telescope at ESO 
La Silla observatory. We tried to reproduce this value by using it as a Gaussian prior (with a width equal the error bar on the 
measurement) in our MCMC. Even with this prior, the resulting occultation depth is $466\pm{97}$ ppm. We are thus not confirming 
the measurement of \citet{Burton12} but conclude that the occultation of WASP-19b in z'~band is significantly smaller. 

In our global analysis we include the occultation observed with HAWK-I using the narrow-band NB1190 filter. 
From our data we measure the occultation depth at 1.19 {\mym} to be $1711_{-730}^{+750}$~ppm.

\section{Discussion}
\label{sec:dis}

From the homogeneous set of TRAPPIST I+z' lightcurves, we can evaluate the photometric improvement obtained from 
combining lightcurves. As presented in Figure \ref{fig:STD}, we are not far from the ideal case of only white noise. 
Even if combining as many as nine lightcurves, we are still gaining in photometric precision by adding additional
transits. 

Following our observation strategy of deriving the most accurate measurement of the overall
transit shape and thus the planetary parameters, we can reduce correlated noise, which can drastically affect 
single lightcurves. This is most evident if the same transit is observed simultaneously using different instruments.
In the case of a combination of nine TRAPPIST lightcurves, we measured the planet/star radius ratio with a precision
of 0.7\% in  I+z'-band, while three observations with EulerCam in r'-band and a combination of two EulerCam lightcurves and
an FTS yield slightly larger errors, giving a precision of 0.9 and 1.0 \%. While these values agree well within their error bars,
a single lightcurve obtained in IC-band gives a 2.6 $\sigma$ lower value. This lightcurve shows a small 
flux increase during transit (possibly the signature of a star spot), and we suggest that this possible radius variation be verified with
additional data. Discarding this point, we see a flat optical transmission spectrum of WASP-19b. Overall, the values derived from our
analysis are in good agreement with the values previously derived.

With two new occultation measurements, we can proceed to constrain the chemical composition and structure of the planetary
atmosphere. For this purpose, we use the model spectra calculated by \citet{Madhu12a}. In their work, they
show models of WASP-19b for a carbon-dominated C/O = 1.1 and an oxygen-dominated C/O = 0.4
atmosphere. For oxygen-rich models, a strong absorption around 0.9 {\mym} is expected from the TiO and VO leading
to a smaller z'-~band occultation depth. Following the occultation measurement by \citet{Burton12}, \citet{Madhu12a} 
tentatively classified WASP-19b as having a carbon-dominated atmosphere. 

Comparing our values to these models (see 
Figure \ref{fig:atm}), we find our measurement of the z'-~band matches the oxygen-rich model extremely well, showing 
higher absorption, indicative of higher abundances in TiO and VO. These elements require a higher 
concentration of oxygen in the planetary atmosphere in order to contribute measurably to the planetary spectrum.
The 1.19 {\mym} value matches the oxygen- and carbon-rich models equally well. Thus, we suggest WASP-19b is a highly irradiated oxygen-dominated planet, fitting the
\textit{O2} or upper \textit{O1} Class defined by \citet{Madhu12a}. This should be confirmed by
a joint analysis of the previously published data and the measurements added in this work since the models of planetary emission are not
unique. 

From the Spitzer data on WASP-19b \citep{Anderson11b}, we know that
WASP-19b does not show any temperature inversion. At first glance, this might seem unexpected for an oxygen-dominated atmosphere,
because it is precisely the molecules of which we measure higher concentrations that are presumed to be causing temperature 
inversions. Still, as different wavelengths probe different depths in the planetary atmosphere,
the z'-~band observations probe a deeper atmospheric layer, which is below the expected temperature inversion.
In this context it would be interesting to evaluate the effects of gravitational settling and stellar UV radiation
on the presence and distribution of TiO and VO in the planetary atmosphere. TiO and VO might be destroyed or depleted from 
the upper atmosphere, thus inhibiting a temperature inversion, while lower in the atmosphere the intact TiO and VO 
could be causing the measured absorption in the z'~band.

\section{Conclusion}
\label{sec:con}

We have carried out an in-depth observing campaign on WASP-19 collecting a total of 14 transit and 10 occultation lightcurves with
EulerCam and TRAPPIST, as well as one 1.19 {\mym} lightcurve with \mbox{HAWK-I}. 
From the large homogeneous set of nine TRAPPIST lightcurves, we demonstrate how both the attainable photometric precision 
and the accuracy of the derived parameters can be greatly improved by combining an increasing number of lightcurves.

We have detected the z'-band occultation of WASP-19b using 1m class telescopes by the combined analysis of our lightcurves. 
We measure it at $352\pm116$~ppm, more than a factor of two smaller than previously published. 
From our HAWK-I data we obtain an occultation depth of $1711_{-730}^{+750}$~ppm at 1.19 {\mym}. These results shed new light on the chemical 
composition of the planetary atmosphere, indicating a $C/O$ ratio of $C/O<1$, i.e. an oxygen-dominated atmosphere.

\begin{acknowledgements}

We would like to thank Nikku Madhusudhan for sharing with us the model spectra of WASP-19b. TRAPPIST is a
project funded by the Belgian Fund for Scientific Research (Fond National de la
Recherche Scientiﬁque, F.R.SFNRS) under grant FRFC 2.5.594.09.F, with the
participation of the Swiss National Science Foundation (SNF). M. Gillon and
E. Jehin are FNRS Research Associates. 

\end{acknowledgements}

\bibliographystyle{aa}
\bibliography{bbl}

\end{document}